\input epsf
\input colordvi
\input amssym
\catcode`@=11                                   
\catcode`\|=12                                  
\catcode`\&=4                                   

\newcount\ncols         \ncols=\z@              
\newcount\nrows         \nrows=\z@              
\newcount\curcol        \curcol=\z@             
     
\newdimen\thinsize      \thinsize=0.6pt         
\newdimen\thicksize     \thicksize=1.5pt        

\newif\iftableinfo      \tableinfotrue          
\newif\ifcentertables   \centertablestrue       
%
%
     
\let\plaincr=\cr                        
\let\plainspan=\span                    
\let\plaintab=&                         
\let\lparen=(                           
\let\NX=\noexpand                       

     
\def\ruledtable{\relax                          
    \@BeginRuledTable                           
    \@RuledTable}


\def\@BeginRuledTable{
   \ncols=0\nrows=0                             
   \begingroup                                  
    \offinterlineskip                           
    \def~{\phantom{0}}
    \def\span{\plainspan\omit\relax\colcount\plainspan}
    \let\cr=\crrule                             
    \let\CR=\crthick                            
    \let\nr=\crnorule                           
    \let\|=\Vb                                  
%
%
    \ifx\tablestrut\undefined\relax             
    \else\let\tstrut=\tablestrut\fi             
    \catcode`\|=13 \catcode`\&=13\relax         
    \TableActive                                
    \curcol=1                                   
%
%
    \ifdim\tablewidth>-\maxdimen\relax          %
      \edef\@Halign{\NX\halign to \NX\tablewidth\NX\bgroup\TablePreamble}%
      \tabskip=0pt plus 1fil                    
    \else                                       %
      \edef\@Halign{\NX\halign\NX\bgroup\TablePreamble}%
      \tabskip=0pt                              
    \fi                                         %
%
%
    \ifcentertables                             
       \ifhmode\vskip 0pt\fi                    
       \line\bgroup\hss                         
    \else\hbox\bgroup                           
    \fi}


\long\def\@RuledTable#1\endruledtable{
   \vrule width\thicksize                       
     \vbox{\@Halign                             
       \thickrule                               
       #1\relax                                 
       \tstrut                                  
       \plaincr\thickrule                       
     \egroup}
   \vrule width\thicksize                       
   \ifcentertables\hss\fi\egroup                
  \endgroup                                     
  \global\tablewidth=-\maxdimen                 
  \iftableinfo                                  
      \immediate\write16{[Nrows=\the\nrows, Ncols=\the\ncols]}%
   \fi}
     

\def\TablePreamble{
   \linecount                           
   \TableItem{####}
   \plaintab\plaintab                   
   \TableItem{####}
   \plaincr}


\def\@TableItem#1{
   \hfil\tablespace                             
   #1\relax                                     
   \tablespace\hfil                             
    }%

\def\@tableright#1{
   \hfil\tablespace\relax               
   #1\relax                             
   \tablespace\relax}

\def\@tableleft#1{
   \tablespace\relax                    
   #1\relax                             
   \tablespace\hfil}

\let\TableItem=\@TableItem              
     
\def\RightJustifyTables{\let\TableItem=\@tableright}
\def\LeftJustifyTables{\let\TableItem=\@tableleft}
\def\NoJustifyTables{\let\TableItem=\@TableItem}

\def\LooseTables{\let\tablespace=\quad}
\def\TightTables{\let\tablespace=\space}
\LooseTables                                    

%

\newdimen\tablewidth    \tablewidth=-\maxdimen  


\def\setRuledStrut{
   \dimen@=\baselineskip                        
   \advance\dimen@ by-\normalbaselineskip       
   \ifdim\dimen@<.5ex \dimen@=.5ex\fi           
   \setbox0=\hbox{\lparen}
   \dimen1=\dimen@ \advance\dimen1 by \ht0      
   \dimen2=\dimen@ \advance\dimen2 by \dp0      
   \def\tstrut{\vrule height\dimen1 depth\dimen2 width\z@}%
   }%

\def\tstrut{\vrule height 3.1ex depth 1.2ex width 0pt}


\def\bigitem#1{
   \setbox0=\hbox{#1}
   \dimen1 =\ht0 \dimen2 =\dp0                  
   \dimen@ =\baselines@ve                       
   \advance\dimen@ by-\normalbaselineskip       
   \ifdim\dimen@<.25ex \dimen@=.25ex\fi         
   \advance\dimen1 by \dimen@                   
   \advance\dimen2 by \dimen@                   
   \vrule height\dimen1 depth\dimen2 width\z@   
   \copy0}

     
%

     
\def\nextcolumn#1{
   \plaintab\omit#1\relax\colcount              
   \plaintab}
     
\def\tab{
   \nextcolumn{\relax}}


\def\vb{
   \nextcolumn{\vrule width\thinsize}}

\def\Vb{
   \nextcolumn{\vrule width\thicksize}}


     
{\catcode`\|=13 \let|0
 \catcode`\&=13 \let&0
 \gdef\TableActive{\let|=\vb \let&=\tab}%
}


\def\crrule{\relax                      
   \tstrut                              
   \plaincr\tablerule                   
  }%

\def\crthick{\relax                     
   \tstrut                              
   \plaincr\thickrule                   
  }%
     
\def\crnorule{\relax                    
   \tstrut                              
   \plaincr                             
   }%
   

     
\def\tablerule{\noalign{\hrule height\thinsize depth 0pt}}%
\def\thickrule{\noalign{\hrule height\thicksize depth 0pt}}%


%
%
%
     

\def\linecount{\relax\global\ncols=\curcol      
   \global\curcol=1                             
   \global\advance\nrows by 1\relax}
     
\def\colcount{\relax                            %
   \global\advance\curcol by 1\relax}


\newdimen\parasize      \parasize=4in           

%

%

\def\begintable{\relax                          
    \@BeginRuledTable                           
    \@begintable}

\long\def\@begintable#1\endtable{
   \@RuledTable#1\endruledtable}


\catcode`@=12                                   


\newfam\scrfam
\batchmode\font\tenscr=rsfs10 \errorstopmode
\ifx\tenscr\nullfont
        \message{rsfs script font not available. Replacing with calligraphic.}
        \def\scr{\cal}
\else   
        \font\sevenscr=rsfs7
        \font\fivescr=rsfs5
        \skewchar\tenscr='177 \skewchar\sevenscr='177 \skewchar\fivescr='177
        \textfont\scrfam=\tenscr \scriptfont\scrfam=\sevenscr
        \scriptscriptfont\scrfam=\fivescr
        \def\scr{\fam\scrfam}
        \def\cal{\scr}
\fi
\catcode`\@=11
\newfam\frakfam
\batchmode\font\tenfrak=eufm10 \errorstopmode
\ifx\tenfrak\nullfont
        \message{eufm font not available. Replacing with italic.}
        
\else
	
	\font\sevenfrak=eufm7 \font\fivefrak=eufm5
	\textfont\frakfam=\tenfrak
	\scriptfont\frakfam=\sevenfrak \scriptscriptfont\frakfam=\fivefrak
	
\fi
\catcode`\@=\active
\newfam\msbfam
\batchmode\font\twelvemsb=msbm10 scaled\magstep1 \errorstopmode
\ifx\twelvemsb\nullfont\def\Bbb{\bf}

	\message{Blackboard bold not available. Replacing with boldface.}
\else   \catcode`\@=11
        \font\tenmsb=msbm10 \font\sevenmsb=msbm7 \font\fivemsb=msbm5
        \textfont\msbfam=\tenmsb
        \scriptfont\msbfam=\sevenmsb \scriptscriptfont\msbfam=\fivemsb
        \def\Bbb{\relax\expandafter\Bbb@}
        \def\Bbb@#1{{\Bbb@@{#1}}}
        \def\Bbb@@#1{\fam\msbfam\relax#1}
        \catcode`\@=\active

\fi
        \font\eightrm=cmr8              \def\xrm{\eightrm}
        \font\eightbf=cmbx8             \def\xbf{\eightbf}
        \font\eightit=cmti10 at 8pt     \def\xit{\eightit}
        
                     
        \font\eightcp=cmcsc8
        \font\eighti=cmmi8              \def\xold{\eighti}
        \font\eightib=cmmib8             \def\xbold{\eightib}
        \font\teni=cmmi10               \def\old{\teni}
        \font\tencp=cmcsc10

        \font\twelvecp=cmcsc10 scaled\magstep1

	 at10pt	
	\font\twelvehelvbold=phvb at12pt
	 at14pt
	\font\sixteenhelvbold=phvb at16pt

\def\noblackbox{\overfullrule=0pt}
\noblackbox

\newtoks\headtext
\headline={\ifnum\pageno=1\hfill\else
	\ifodd\pageno{\eightcp\the\headtext}{ }\dotfill{ }{\old\folio}
	\else{\old\folio}{ }\dotfill{ }{\eightcp\the\headtext}\fi
	\fi}
\def\makeheadline{\vbox to 0pt{\vss\noindent\the\headline\break
\hbox to\hsize{\hfill}}
        \vskip2\baselineskip}
\newcount\infootnote
\infootnote=0
\def\foot#1#2{\infootnote=1
\footnote{${}^{#1}$}{\vtop{\baselineskip=.75\baselineskip
\advance\hsize by -\parindent\noindent{\xrm #2}}}\infootnote=0$\,$}
\newcount\refcount
\refcount=1
\newwrite\refwrite
\def\oldsize{\ifnum\infootnote=1\xold\else\old\fi}
\def\ref#1#2{
	\def#1{{{\oldsize\the\refcount}}\ifnum\the\refcount=1\immediate\openout\refwrite=\jobname.refs\fi\immediate\write\refwrite{\item{[{\xold\the\refcount}]} 
	#2\hfill\par\vskip-2pt}\xdef#1{{\noexpand\oldsize\the\refcount}}\global\advance\refcount by 1}
	}
\def\refout{\catcode`\@=11
        \xrm\immediate\closeout\refwrite
        \vskip2\baselineskip
        {\noindent\twelvecp References}\hfill\vskip\baselineskip
        \baselineskip=.75\baselineskip
        \input\jobname.refs
        \baselineskip=4\baselineskip \divide\baselineskip by 3
        \catcode`\@=\active\rm}

\def\hepth#1{\href{http://xxx.lanl.gov/abs/hep-th/#1}{hep-th/{\xold#1}}}

\def\arxiv#1#2{\href{http://arxiv.org/abs/#1.#2}{arXiv:{\xold#1}.{\xold#2}}}
\def\jhep#1#2#3#4{\href{http://jhep.sissa.it/stdsearch?paper=#2\%28#3\%29#4}{J. High Energy Phys. {\xbold #1#2} ({\xold#3}) {\xold#4}}}
\def\AP#1#2#3{Ann. Phys. {\xbold#1} ({\xold#2}) {\xold#3}}

\def\CQG#1#2#3{Class. Quantum Grav. {\xbold#1} ({\xold#2}) {\xold#3}}

\def\JHEP{\jhep}

\def\JPA#1#2#3{J. Phys. {\xbf A}{\xbold#1} ({\xold#2}) {\xold#3}}

\def\MPLA#1#2#3{Mod. Phys. Lett. {\xbf A}{\xbold#1} ({\xold#2}) {\xold#3}}

\def\NPB#1#2#3{Nucl. Phys. {\xbf B}{\xbold#1} ({\xold#2}) {\xold#3}}

\def\PLB#1#2#3{Phys. Lett. {\xbf B}{\xbold#1} ({\xold#2}) {\xold#3}}

\def\PRD#1#2#3{Phys. Rev. {\xbf D}{\xbold#1} ({\xold#2}) {\xold#3}}
\def\PRL#1#2#3{Phys. Rev. Lett. {\xbold#1} ({\xold#2}) {\xold#3}}

\newcount\sectioncount
\sectioncount=0
\def\section#1#2{\global\eqcount=0
	\global\subsectioncount=0
        \global\advance\sectioncount by 1
	\ifnum\sectioncount>1
	        \vskip2\baselineskip
	\fi
\line{\twelvecp\the\sectioncount. #2\hfill}
       \vskip.5\baselineskip\noindent
        \xdef#1{{\old\the\sectioncount}}}
\newcount\subsectioncount
\def\subsection#1#2{\global\advance\subsectioncount by 1
	\vskip.75\baselineskip\noindent
\line{\tencp\the\sectioncount.\the\subsectioncount. #2\hfill}
	\vskip.5\baselineskip\noindent
	\xdef#1{{\old\the\sectioncount}.{\old\the\subsectioncount}}}
\def\immediatesubsection#1#2{\global\advance\subsectioncount by 1
\vskip-\baselineskip\noindent
\line{\tencp\the\sectioncount.\the\subsectioncount. #2\hfill}
	\vskip.5\baselineskip\noindent
	\xdef#1{{\old\the\sectioncount}.{\old\the\subsectioncount}}}
\newcount\appendixcount
\appendixcount=0
\def\appendix#1{\global\eqcount=0
        \global\advance\appendixcount by 1
        \vskip2\baselineskip\noindent
        \ifnum\the\appendixcount=1
        \hbox{\twelvecp Appendix A: #1\hfill}\vskip\baselineskip\noindent\fi
    \ifnum\the\appendixcount=2
        \hbox{\twelvecp Appendix B: #1\hfill}\vskip\baselineskip\noindent\fi
    \ifnum\the\appendixcount=3
        \hbox{\twelvecp Appendix C: #1\hfill}\vskip\baselineskip\noindent\fi}

\newcount\eqcount
\eqcount=0
\def\Eqn#1{\global\advance\eqcount by 1
\ifnum\the\sectioncount=0
	\xdef#1{{\old\the\eqcount}}
	\eqno({\oldstyle\the\eqcount})
\else
        \ifnum\the\appendixcount=0
	        \xdef#1{{\old\the\sectioncount}.{\old\the\eqcount}}
                \eqno({\oldstyle\the\sectioncount}.{\oldstyle\the\eqcount})\fi
        \ifnum\the\appendixcount=1
	        \xdef#1{{\oldstyle A}.{\old\the\eqcount}}
                \eqno({\oldstyle A}.{\oldstyle\the\eqcount})\fi
        \ifnum\the\appendixcount=2
	        \xdef#1{{\oldstyle B}.{\old\the\eqcount}}
                \eqno({\oldstyle B}.{\oldstyle\the\eqcount})\fi
        \ifnum\the\appendixcount=3
	        \xdef#1{{\oldstyle C}.{\old\the\eqcount}}
                \eqno({\oldstyle C}.{\oldstyle\the\eqcount})\fi
\fi}
\def\eqn{\global\advance\eqcount by 1
\ifnum\the\sectioncount=0
	\eqno({\oldstyle\the\eqcount})
\else
        \ifnum\the\appendixcount=0
                \eqno({\oldstyle\the\sectioncount}.{\oldstyle\the\eqcount})\fi
        \ifnum\the\appendixcount=1
                \eqno({\oldstyle A}.{\oldstyle\the\eqcount})\fi
        \ifnum\the\appendixcount=2
                \eqno({\oldstyle B}.{\oldstyle\the\eqcount})\fi
        \ifnum\the\appendixcount=3
                \eqno({\oldstyle C}.{\oldstyle\the\eqcount})\fi
\fi}
\def\multi{\global\advance\eqcount by 1}
\def\multieq#1#2{\xdef#1{{\old\the\eqcount#2}}
        \eqno{({\oldstyle\the\eqcount#2})}}
\newtoks\url
\def\Href#1#2{\catcode`\#=12\url={#1}\catcode`\#=\active#2}
\def\href#1#2{{#2}}

\parskip=3.5pt plus .3pt minus .3pt
\baselineskip=14pt plus .1pt minus .05pt
\lineskip=.5pt plus .05pt minus .05pt
\lineskiplimit=.5pt
\abovedisplayskip=18pt plus 4pt minus 2pt
\belowdisplayskip=\abovedisplayskip
\hsize=14cm
\vsize=19cm
\hoffset=1.5cm
\voffset=1.8cm
\frenchspacing
\footline={}
\raggedbottom

\def\ss{\scriptstyle}
\def\sss{\scriptscriptstyle}
\def\*{\partial}
\def\punkt{\,\,.}
\def\komma{\,\,,}

\def\={\!=\!}
\def\small#1{{\hbox{$#1$}}}

\def\fraction#1{\small{1\over#1}}
\def\fr{\fraction}
\def\Fraction#1#2{\small{#1\over#2}}
\def\Fr{\Fraction}
\def\tr{\hbox{\rm tr}}
\def\eg{{\tenit e.g.}}

\def\ie{{\tenit i.e.}}
\def\etal{{\tenit et al.}}

\def\a{\alpha}
\def\b{\beta}

\def\d{\delta}
\def\e{\varepsilon}
\def\g{\gamma}
\def\l{\lambda}

\def\w{\!\wedge\!}
\def\id{1\hskip-3.5pt 1}

\def\Dslash{D\hskip-6.5pt/\hskip1.5pt}


\def\modprod#1{\raise0pt\vtop{\baselineskip=0pt\lineskip=0pt
      \ialign{\hfill##\hfill\cr$\circ$\cr${\sss #1}$\cr}}}

\def\third{3\raise3pt\hbox{\eightit rd}}
\def\fourth{4\raise3pt\hbox{\eightit th}}


\def\monthname{\ifcase\month 0/\or January\or February\or
   March\or April\or May\or June\or July\or August\or September\or
   October\or November\or December
\fi}


\def\boxit#1{\vbox{\hsize=13cm\hrule\hbox{\vrule\kern6pt
             \vbox{\kern3pt\noindent#1\hfill\break\kern6pt}
             \kern6pt\vrule\hfill}\hrule}}



\ref\KuzenkoLindstromTartaglino{S.M. Kuzenko, U. Lindstr\"om and
G. Tartaglino-Mazzucchelli, {\xit ``Off-shell supergravity--matter
couplings in three dimensions''}, \arxiv{1101}{4013}.}

\ref\HoweIzquierdoPapadopoulosTownsend{P. Howe, J.M. Izquierdo,
G. Papadopoulos and P.K. Townsend, {\xit ``New supergravities with
central charges and Killing spinors in 2+1
dimensions''}, \NPB{467}{1996}{183} [\hepth{9505032}].}

\ref\HoweWeyl{P.~Howe,
{\xit ``Weyl superspace''},
\PLB{415}{1997}{149} [\hepth{9707184}].}

\ref\HoweExtended{P. Howe, {\xit ``A superspace approach to extended
conformal supergravity''}, \PLB{100}{1981}{389}.}

\ref\GranNilsson{U. Gran and B.E.W. Nilsson, {\xit ``Three-dimensional
N=8 superconformal gravity and its coupling to BLG
M2-branes''}, \jhep{09}{03}{2009}{074} [\arxiv{0809}{4478}].}

\ref\ConvConstr{S.J.~Gates, K.S.~Stelle and P.C.~West,
{\xit ``Algebraic origins of superspace constraints in supergravity''},
\NPB{169}{1980}{347}; 
S.J. Gates and W. Siegel, 
{\xit ``Understanding constraints in superspace formulation of supergravity''},
\NPB{163}{1980}{519}.}

\ref\CGNT{M. Cederwall, U. Gran, B.E.W. Nilsson and D. Tsimpis,
{\xit ``Supersymmetric corrections to eleven-dimen\-sional supergravity''},
\jhep{05}{05}{2005}{052} [\hepth{0409107}].}

\ref\CederwallBLG{M. Cederwall, {\xit ``N=8 superfield formulation of
the Bagger--Lambert--Gustavsson model''}, \jhep{08}{09}{2008}{116}
[\arxiv{0808}{3242}].}

\ref\CederwallABJM{M. Cederwall, {\xit ``Superfield actions for N=8 
and N=6 conformal theories in three dimensions''},
\jhep{08}{10}{2008}{70}
[\arxiv{0808}{3242}].}

\ref\Dragon{N. Dragon, {\xit ``Torsion and curvature in extended
supergravity''}, Z. Phys. {\xbf C2} (1979) 29.}

\ref\BaggerLambertI{J. Bagger and N. Lambert, {\xit ``Modeling
multiple M2's''}, \PRD{75}{2007}{045020} [\hepth{0611108}].}

\ref\BaggerLambertII{J. Bagger and N. Lambert, {\xit ``Gauge symmetry
and supersymmetry of multiple M2-branes''}, \PRD{77}{2008}{065008}
[\arxiv{0711}{0955}].} 

\ref\Gustavsson{A. Gustavsson, {\xit ``Algebraic structures on
parallel M2-branes''}, \arxiv{0709}{1260}.}

\ref\MukhiPapa{S. Mukhi and C. Papageorgakis, {\xit ``M2 to
D2''}, \jhep{08}{05}{2008}{085} [\arxiv{0803}{3218}].}

\ref\BST{E. Bergshoeff, E. Sezgin and P.K. Townsend,
	``Supermembranes and eleven-dimensional supergravity'',
	\PLB{189}{1987}{75}; 
	``Properties of the eleven-dimensional super membrane theory'',
	\AP{185}{1988}{330}.}

\ref\LiSongStrominger{W. Li, W. Song and A. Strominger, {\xit ``Chiral
	gravity in three dimensions''}, \jhep{08}{04}{2008}{082}
	[\arxiv{0801}{4566}].}

\ref\CGNN{M. Cederwall, U. Gran, M. Nielsen and B.E.W. Nilsson, 
{\xit ``Manifestly supersymmetric M-theory''}, 
\JHEP{00}{10}{2000}{041} [\hepth{0007035}];
{\xit ``Generalised 11-dimensional supergravity''}, \hepth{0010042}.}

\ref\ABJM{O. Aharony, O. Bergman, D.L. Jafferis and J. Maldacena,
{\xit ``N=6 superconformal Chern--Simons-matter theories, M2-branes
and their gravity duals''}, \arxiv{0806}{1218}.}

\ref\GreitzHowe{J. Greitz and P.S. Howe, {\xit ``Maximal supergravity
in three dimensions: supergeometry and differential
forms''}, \arxiv{1103}{2730}.} 

\ref\BerkovitsIII{N. Berkovits, 
{\xit ``Cohomology in the pure spinor formalism for the
superstring''}, 
\jhep{00}{09}{2000}{046} [\hepth{0006003}].}

\ref\BerkovitsParticle{N. Berkovits, {\xit ``Covariant quantization of
the superparticle using pure spinors''}, \jhep{01}{09}{2001}{016}
[\hepth{0105050}].}

\ref\CederwallNilssonTsimpis{M. Cederwall, B.E.W. Nilsson and D. Tsimpis, 
{\xit ``The structure of maximally supersymmetric super-Yang--Mills theory---constraining higher order corrections''}, \jhep{01}{06}{2001}{034} 
[\hepth{0102009}].}

\ref\SpinorialCohomology{M. Cederwall, B.E.W. Nilsson and D. Tsimpis, 
{\xit ``Spinorial cohomology and maximally supersymmetric theories''},
\jhep{02}{02}{2002}{009} [\hepth{0110069}];
M. Cederwall, {\xit ``Superspace methods in string theory,
supergravity and gauge theory''}, Lectures at the XXXVII Winter School
in Theoretical Physics ``New Developments in Fundamental Interactions
Theories'',  Karpacz, Poland,  Feb. 6-15, 2001, \hepth{0105176}.}

\ref\DuffHoweInamiStelle{M.J. Duff, P.S. Howe, T. Inami and K.S. Stelle,
{\xit ``Superstrings in D=10 from supermembranes in D=11''},
\PLB{191}{1987}{70}.}

\ref\LindstromRocek{U. Lindstr\"om and M. Ro\v cek, {\xit
``Superconformal gravity in three-dimensions as a gauge
theory''}, \PRL{62}{1989}{2905}.} 

\ref\DeserKay{S. Deser and J.H. Kay, {\xit ``Topologically massive
	supergravity''}, \PLB{120}{1983}{97}.}

\ref\VanNCS{P. van Nieuwenhuizen, {\xit ``D=3 conformal supergravity
 and Chern--Simons terms''}, \PRD{32}{1985}{872}.}

\ref\ChuNilsson{X. Chu and B.E.W. Nilsson, {\xit ``Three-dimensional
	topologically gauged N=6 ABJM type theories''}, 
\jhep{10}{06}{2010}{057} [\arxiv{0906}{1655}].}

\ref\CNNP{X. Chu, H. Nastase, B.E.W. Nilsson and C. Papageorgakis,
{\xit ``Higgsing M2 to D2 with gravity: N=6 chiral supergravity from
topologically gauged ABJM theory''}, \arxiv{1012}{5969}.}

\ref\NilssonTollsten{B.E.W. Nilsson and A.K. Tollst\'en, {\xit ``The
geometrical off-shell structure of pure N=1 D=10 supergravity in
superspace''}, \PLB{169}{1986}{369}.}

\ref\BermanPerry{D.S. Berman and M.J. Perry, {\xit ``M-theory and the
    string genus expansion''}, \PLB{635}{2006}{131} [\hepth{0601141}].}

\ref\HoweUmerski{P.S. Howe and A. Umerski, {\xit ``On superspace
supergravity in ten dimensions''}, \PLB{177}{1986}{163}.}

\ref\HoweTwoDim{P.S. Howe, {\xit ``Super Weyl transformations in
two-dimensions}, \JPA{12}{1979}{393}.}

\ref\PureSG{M. Cederwall, {\xit ``Towards a manifestly supersymmetric
    action for D=11 supergravity''}, \jhep{10}{01}{2010}{117}
    [\arxiv{0912}{1814}],  
{\xit ``D=11 supergravity with manifest supersymmetry''},
    \MPLA{25}{2010}{3201} [\arxiv{1001}{0112}].}

\ref\BNXXXI{B.E.W.~Nilsson, 
{\xit ``Off-shell d= 10, N=1 Poincar\'e  supergravity and the embeddibility
of higher derivative field theories in superspace''},
\PLB{175}{1986}{319}.}

\ref\HoweSezginRevisited{P.S. Howe and E. Sezgin, {\xit ``The
supermembrane revisited''}, \CQG{22}{2005}{2167} [\hepth{0412245}].}

\ref\NilssonPure{B.E.W.~Nilsson,
{\xit ``Pure spinors as auxiliary fields in the ten-dimensional
supersymmetric Yang--Mills theory''},
\CQG3{1986}{{\xrm L}41}.}

\ref\HowePureI{P.S. Howe, {\xit ``Pure spinor lines in superspace and
ten-dimensional supersymmetric theories''}, \PLB{258}{1991}{141}.}

\ref\HowePureII{P.S. Howe, {\xit ``Pure spinors, function superspaces
and supergravity theories in ten and eleven dimensions''},
\PLB{273}{1991}{90}.}


\headtext={Cederwall, Gran, Nilsson: 
``D=3, N=8 conformal supergravity and the Dragon window''}

\line{
\epsfxsize=18mm
\epsffile{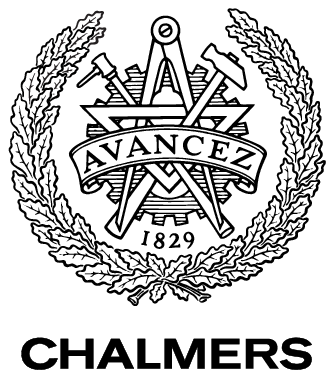}
\hfill}
\vskip-12mm
\line{\hfill G\"oteborg preprint}
\line{\hfill March, {\old2011}; revised August {\old2011}}
\line{\hrulefill}

\vfill
\vskip.5cm

\centerline{\sixteenhelvbold
D = 3, N = 8 conformal supergravity}

\vskip2\parskip
\centerline{\sixteenhelvbold
and the Dragon window} 

\vskip4\parskip


\vfill

\centerline{\twelvehelvbold
Martin Cederwall, Ulf Gran and Bengt E.W. Nilsson}

\vfill

\centerline{\it Fundamental Physics}
\centerline{\it Chalmers University of Technology}
\centerline{\it SE 412 96 G\"oteborg, Sweden}

\vfill

{\narrower\noindent\underbar{Abstract:} We give a superspace
description of $D=3$, $N=8$ supergravity. The formulation is off-shell
in the sense that the equations of motion are not implied by the
superspace constraints (but an action principle is not given). The
multiplet structure is unconventional, which we connect to the
existence of a ``Dragon window'', that is modules occurring in the
supercurvature but not in the supertorsion. 
According to Dragon's theorem this cannot happen
above three dimensions. We clarify the relevance
of this window for going on the conformal shell, and discuss some
aspects of coupling to conformal matter.
\smallskip}
\vfill

\font\xxtt=cmtt6

\vtop{\baselineskip=.6\baselineskip\xxtt
\line{\hrulefill}
\catcode`\@=11
\line{email: martin.cederwall@chalmers.se, ulf.gran@chalmers.se, tfebn@chalmers.se\hfill}
\catcode`\@=\active
}

\eject

\def\l{\lambda}

\def\tZ{\tilde Z}
\def\tz{\tilde z}

\def\textfrac#1#2{\leavevmode
\kern-.1em\raise .45ex \hbox{\the\scriptfont0 #1}
\kern-.50em $/$
\kern-.48em \lower .20ex \hbox{\the\scriptfont0 #2}
\kern-.28em}

\def\textfrac#1#2{\raise .45ex\hbox{\the\scriptfont0 #1}\nobreak\hskip-1pt/\hskip-1pt\hbox{\the\scriptfont0 #2}}

\def\tf#1#2{\textfrac#1#2}

\def\ms{{\mathstrut}}

\def\A{{\cal A}}
\def\B{{\cal B}}


\section\Introduction{Introduction}The recent discovery in M-theory of
Lagrangian formulations for stacks 
of M2-branes at the IR fix-point has lead to an enormous boost in the
interest in three-dimensional superconformal Chern--Simons matter
theories. This development was triggered by the construction of a
theory  
with eight conformal supersymmetries, related to a stack with two
M2-branes, by Bagger and Lambert [\BaggerLambertI,\BaggerLambertII], 
and independently by Gustavsson [\Gustavsson]
(BLG). This was soon generalised by Aharony \etal\ [\ABJM]
 to an arbitrary number of M2-branes in terms of a Lagrangian
 exhibiting, however, only six supersymmetries (ABJM). 

One may ask how these theories are related to string theory in ten
dimensions, that is to fundamental strings and D2-branes.  
Based on a new kind of Higgs mechanism, a connection of the BLG theory to 
D2-branes was found by Mukhi and Papageorgakis in ref. [\MukhiPapa],
that turns the BLG $SU(2)\times SU(2)$ Chern--Simons theory into an ordinary
$SU(2)$  Yang--Mills theory appropriate for a 2-stack of
D2-branes. This mechanism has subsequently been applied to ABJM and a
number of other theories, see ref. [\CNNP] and the references therein.  

However, in the case of the fundamental string the situation is
different. While it is well understood
[\DuffHoweInamiStelle,\BermanPerry] 
how to relate the Bergshoeff \etal\ [\BST] 
non-conformal theory for one M2-brane 
to the string, much  less is known about how 
the  Polyakov formulation of the string is related to the
superconformal BLG theory.  One might suspect that to make such a
relation explicit also the M2-branes should have a version involving
non-dynamical supergravity. Given the Lagrangian for extended
conformal supergravity  
for any number of supersymmetries [\DeserKay,\VanNCS,\LindstromRocek], 
constructions of this type, referred to as
topologically gauged matter theories in the following references,
were obtained  
in the BLG case in ref. [\GranNilsson], where $N=8$ conformal supergravity
was coupled to the BLG theory and the
Lagrangian partly  
derived, and in the ABJM case in ref. [\ChuNilsson], where the coupled
Lagrangian was obtained in full detail. A curious result pointed out
in the latter work is that if this  topologically gauged ABJM theory
is higgsed one finds that the theory ends a up at a chiral point
similar
to the one of Li, Song and Strominger [\LiSongStrominger]. This connection to
D2-branes at the chiral point was further discussed in ref. [\CNNP]
where many of the  details were worked out. 

The aim of this paper is to find a superspace formulation of
three-dimensional supergravity  that can  accommodate not only
Poincar\'e supergravity but also conformal supergravity 
together with the currents needed for the matter couplings in the
topologically gauged BLG theory [\GranNilsson]. The natural starting
point is the standard set of superspace Bianchi identities (BI's) for
the supertorsion and supercurvature. Prior to imposing any
constraints on the fields, they describe "off-shell" supergravity 
in the sense that no
dynamical equations are hidden in the BI's. The task is thus
to find constraints that lead to dynamical equations 
of the required kind. If these equations can be obtained containing
arbitrary currents then we say that the theory is still
off-shell. This usage of the concept "off-shell" has been adopted in
other circumstances as well. For a more elaborate explanation in the
context of eleven-dimensional supergravity, see ref. [\CGNN]. In general
these methods are well suited for the derivation of Poincar\'e
supergravity theories in eleven dimensions  and below, and for any
number of supersymmetries.  

For conformal supergravities these superspace methods have been
 much less studied. Although related issues
have been discussed both in the
past 
[\HoweIzquierdoPapadopoulosTownsend,\HoweExtended,\HoweWeyl,\NilssonTollsten,\HoweSezginRevisited] 
and more  recently [\KuzenkoLindstromTartaglino], 
to our knowledge no field
equations for conformal supergravity have so far been derived with
these methods. In three dimensions one needs to understand  how to
obtain the Cotton equation and its spin \tf32 analog, the Cottino
equation, from the BI's. By essentially extending the analysis in
ref. [\HoweIzquierdoPapadopoulosTownsend], we
will here demonstrate how this can be done by solving the relevant BI
equations to the required level (dimension). This calculation is more
complicated than the corresponding one for Poincar\'e supergravity
simply because the conformal field equations have one extra derivative which
forces one to carry the calculation one level (dimension) higher. We
should remark also that the constraints used to derive the field
equations for Poincar\'e supergravity must be relaxed in the
superconformal case since \eg\ the Ricci tensor is not itself
constrained by the field equations (as in the Poincar\'e case) but
appear in the Cotton equation with an extra derivative acting on it. 
This will be explained in detail below.

Our final results rely on a phenomenon unique to three dimensions
which we call the ``Dragon window''. It refers to the fact that there
are irreducible components in the supercurvature tensor that do not
appear in the supertorsion. In dimensions above three a classic theorem by
Dragon [\Dragon] guarantees that this does not happen, which has the
further consequence that the Bianchi identities for the supercurvature
are automatically satisfied once the Bianchi identities for the
supertorsion are solved. Thus in the case of three dimensions
investigated here  
also independent curvature BI's must be solved. The Dragon window
phenomenon was implicitly found in refs. 
[\HoweIzquierdoPapadopoulosTownsend,\HoweSezginRevisited]\foot\star{We thank Paul Howe for pointing
this out to us.} and
more recently discussed in ref. [\KuzenkoLindstromTartaglino].  

This paper is organised as follows. We continue in section {\old2} with a
discussion of the structure of the supermultiplet  that we expect to
find when solving the BI's. 
This is partly carried out in the language of pure spinor cohomology
which gives  independent insight into the multiplet structure. 
In section {\old3} we first analyse the available constraints and then go on
to solve the BI's both for the torsion and the curvature up to the
level needed to conclude that the conformal supergravity equations,
like the Cotton equation, are all present and fairly easily derivable. The
dynamical equations are further discussed in section {\old4} along with 
some comments and observations about possible matter couplings. Some
final conclusions are collected in section {\old5}.

\vfill\eject
\section\Multiplets{The supergravity multiplet}Let us denote 
3-dimensional vector indices $a,b,\ldots$, and
collective spinor indices $\a A,\b B,\ldots$, where $\a,\b,\ldots=1,2$ are
$so(1,2)\approx sl(2)$ spinor indices and $A,B,\ldots=1,\ldots,8$
chiral $so(8)$ spinor
indices ($8_s$). We also write the spinor index of opposite chirality 
($8_c$) as 
$A',B',\ldots$ and so(8) vectors ($8_v$) as $I,J,\ldots$. 
Collective superspace flat tangent indices are written $\A,\B,\ldots$,
with $\A=(a,\a A)$.
Dynkin labels have
the sl(2) node in the first position. We use the triality convention
$8_v=(1000)$, $8_s=(0010)$, $8_c=(0001)$.
Conventions for
3-dimensional spinors
are that indices are lowered and raised with $\e_{\a\b}$ and its
inverse. 
They are raised and
lowered by matrix multiplication, so that \eg\
$M^\a{}_\b=M^{\a\g}\e_{\g\b}$ and $M_\a{}^\b=\e_{\a\g}M^{\g\b}$. 
Then the sign issues associated with
symplectic spinor modules are kept to a minimum, although one
still has to remember details like $(\g_a)_\a{}^\b=-(\g_a)^\b{}_\a$. We
use conventions where $(\g_{abc})_\a{}^\b=+\e_{abc}\d_\a{}^\b$.

A convenient way to investigate linearised supermultiplets is to use
pure spinor cohomology
[\BerkovitsIII,\BerkovitsParticle,\CederwallNilssonTsimpis,\SpinorialCohomology]
(see also refs. [\NilssonPure,\HowePureI,\HowePureII] for early work
on the r\^ole of pure spinors for supersymmetric theories). 
Since the $N=8$ supergravity multiplet is
half-maximally supersymmetric, we expect to find an off-shell
supermultiplet. We take the pure spinor $\l^{\a A}$ to transform as
$(1)(0010)$ of $sl(2)\oplus so(8)$. An arbitrary symmetric spinor
bilinear is in $(2)(0000)\oplus(0)(0100)\oplus(2)(0020)$. The torsion at 
dimension 0 is $T_{\a A,\b B}{}^c=2\d_{AB}\g^c_{\a\b}$. 
It is straightforward to check (see the following section) 
that conventional constraints
[\ConvConstr,\CGNT] may be used to
set everything else except a component in $(4)(0020)$ to zero in the
dimension 0 torsion. The pure spinor field, whose lowest component is
the diffeomorphism ghost, is of course taken to be 
$\Phi^a$ in $(2)(0000)$, with an extra gauge invariance 
$$
\Phi^a\approx\Phi^a+(\l\g^a\rho)\Eqn\PhiGaugeInv
$$ 
(as is standard in the description
of gravity multiplets, and more generally, for non-scalar pure spinor
superfields). The nilpotency of the BRST operator
$Q=\l^{\a A}D_{\a A}$ implies that the $D=3$
vector $\l^{\a A}\l^{\b A}$ has to vanish.
It is not {\it a priori} clear whether or not also the bilinear in
$(0)(0100)$ will vanish. 
We know from the superspace formulation of
the BLG model [\CederwallBLG,\CederwallABJM] that the scalar multiplet
relies on pure spinors with non-vanishing bilinear in
$(0)(0100)$. 
Here, it is straightforward to show that the pure spinor constraint in
$(2)(0000)$ together with the gauge invariance (\PhiGaugeInv) ensures
that the power expansion in the pure spinor gives the irreducible
module $(n+2)(00n0)$ at order $\l^n$. Therefore, at $\l^2$ the correct
torsion in $(4)(0020)$ is reproduced. The supergravity is
formulated using the same pure spinors as in the BLG model, and there
is no need to constrain the bilinear in $(0)(0100)$.

The zero-mode cohomology looks as follows:
\def\wsp{\phantom{00000000000}}
\def\wssp{\phantom{000}}
\vskip2\parskip
\vbox{
$$
\vtop{\baselineskip20pt\lineskip0pt
\ialign{
$\hfill#\quad$&$\,\hfill#\hfill\,$&$\,\hfill#\hfill\,$&$\,\hfill#\hfill\,$
&$\,\hfill#\hfill\,$&$\,\hfill#\hfill\,$\cr
            &\hbox{gh}\#=1    &0    &-1    &-2  \cr
\hbox{dim}=-1&\wssp(2)(0000)\wssp&\wsp&       &       \cr
        -\Fr12&(1)(0010)&\bullet&\wsp&  \cr
           0&(0)(0100)&(4)(0000)&\bullet&\wsp\cr
       \Fr12&\bullet&(3)(0010)&\bullet&\bullet     \cr
           1&\bullet&(0)(2000)\oplus(2)(0100)\Gray{\oplus(0)(0002)}
                                &\bullet&\bullet\cr
       \Fr32&\bullet&(1)(1001)\Gray{\oplus(1)(1001)}&\bullet&\bullet\cr
           2&\bullet&(0)(0002)\Gray{\oplus(0)(2000)\oplus(2)(0100)}
                                &\bullet&\bullet\cr
       \Fr52&\bullet&\Gray{(3)(0010)}&\bullet&\bullet\cr
           3&\bullet&\Gray{(4)(0000)}&\Gray{(0)(0100)}&\bullet\cr
       \Fr72&\bullet&\bullet&\Gray{(1)(0010)}&\bullet\cr
           4&\bullet&\bullet&\Gray{(2)(0000)}&\bullet\cr
       \Fr92&\bullet&\bullet&\bullet&\bullet\cr
}}
$$
}
\vskip2\parskip
\noindent The structure of the pure spinor cohomology is intriguing, in that it
provides an off-shell formulation, but still contains a ``current
multiplet'' in the same field (the component $E_{\a A}{}^a$ of the
super-vielbein) as the physical fields. In the first column we
recognise the ghosts (or gauge parameters) for superdiffeomorphisms
and local $so(8)$ transformations. In the second column, the off-shell
supergravity multiplet contains the linearised graviton, gravitino and
$so(8)$ gauge potential, as well as auxiliary fields at dimensions 1,
\tf32 and 2. By implementing an extra constraint setting the field in
$(0)(0002)$ at dimension 1 to zero, the fields in grey are
eliminated. 
The constraint implements a kind of selfduality in the multiplet
structure --- there is nothing in the previously introduced data that
distinguishes the two modules $(0)(2000)$ (selfdual in four spinor
indices) and $(0)(0002)$ (anti-selfdual).

The current (or antifield) multiplet can be
obtained in a similar way. It comes in a field in the module
$(0)(2000)$ (there is also a gauge invariance analogous
to that of eq. (\PhiGaugeInv)). This antifield zero-mode cohomology is:
\def\wsp{\phantom{00000000000}}
\def\wssp{\phantom{000}}
\vskip2\parskip
\vbox{
$$
\vtop{\baselineskip20pt\lineskip0pt
\ialign{
$\hfill#\quad$&$\,\hfill#\hfill\,$&$\,\hfill#\hfill\,$&$\,\hfill#\hfill\,$
&$\,\hfill#\hfill\,$&$\,\hfill#\hfill\,$\cr
            &\hbox{gh}\#=-1    &-2    &-3    &-4  \cr
\hbox{dim}=1&\wssp(0)(2000)\wssp&\wsp&       &       \cr
        \Fr32&(1)(1001)&\bullet&\wsp&  \cr
           2&(0)(00002)\oplus(2)(0100)&\bullet&\bullet&\wsp\cr
       \Fr52&(3)(0010)&\bullet&\bullet&\bullet     \cr
           3&(4)(0000)&(0)(0100)&\bullet&\bullet\cr
       \Fr72&\bullet&(1)(0010)&\bullet&\bullet\cr
           4&\bullet&(2)(0000)&\bullet&\bullet\cr
       \Fr92&\bullet&\bullet&\bullet&\bullet\cr
           5&\bullet&\bullet&\bullet&\bullet\cr
}}
$$
}
\noindent Note that this multiplet consists of a triality rotated version of
the modules in grey in the field cohomology (present before the extra
constraint is imposed), 
and that there is room for the Cotton and Cottino equations at
dimensions 3 and \tf52.

The field multiplet has the peculiar feature that supersymmetry
connects the component fields in a tree-like structure:

\epsfysize=2cm
\hskip4cm\epsffile{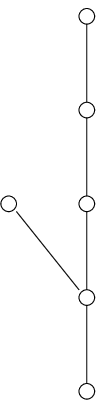}

\noindent This property is read off from the $sl(2)$ modules appearing. The
field at the ``branch'' of the tree is the auxiliary field $C^+_{IJ}$ in
$(0)(2000)$. 
The antifield/current multiplet of course has the same structure, but
turned upside down.
In the following sections we will demonstrate that its
appearance in the superspace geometry is a result of what we choose to
call a ``Dragon window'' --- a module appearing in the supercurvature
but not in the supertorsion. The proof of Dragon's theorem [\Dragon]
assumes the space-time dimension to be greater than 3.
This field plays a central r\^ole for the equations of motion in
conformal supergravity. 

The conformal multiplet for all $N$ were discussed in
ref. [\HoweIzquierdoPapadopoulosTownsend] and more recently in refs. 
[\KuzenkoLindstromTartaglino,\GreitzHowe], based on an investigation of the
superspace Bianchi identities, while the $N=8$ multiplet with the
selfduality constraint on the dimension 1 scalars was given in
refs. [\HoweIzquierdoPapadopoulosTownsend,\HoweSezginRevisited]. There,
the authors presented the proper 
constraints and indicated how the BI's result in the above structure of the
$N=8$ supermultiplet. In the following section we continue this
analysis to the extent that we can conclude that the conformal field
equations are actually derivable in this approach.

\section\BianchiIdentities{Solving the Bianchi identities}In this
section, we will show how the solution of the the superspace Bianchi
identities leads to the off-shell supergravity multiplet.
Torsion is defined as $T^\A=DE^\A=dE^\A+E^\B\wedge\Omega_\B{}^\A$,
where $E^\A$ is the superspace vielbein (frame 1-form) and
$\Omega_\A{}^\B$ the superspace connection 1-form, taking values in
the structure algebra. The torsion Bianchi identity therefore reads 
$DT^\A=E^\B\wedge R_\B{}^\A$. Since, as will be explained below,  
Dragon's theorem does not fully
apply in $D=3$, we also need the curvature Bianchi identity
$DR_\A{}^\B=0$.

The structure algebra would na\"\i vely be
taken as the sum of the Lorentz algebra and the R-symmetry algebra,
but we will also include Weyl scalings, so that the total structure
algebra is $so(1,2)\oplus so(8)\oplus{\Bbb R}$. Although not necessary
for conformal symmetry, the inclusion of Weyl
scalings will facilitate the treatment of scaling properties and the
elimination of unphysical degrees of freedom through conventional
constraints.
An element $T$ in the fundamental representation of the structure algebra 
(\ie, acting on a superspace tangent vector) has the non-vanishing
matrix elements
$$
\eqalign{
T_a{}^b&=L_a{}^b+N\delta_a{}^b\komma\cr
T_{\a A}{}^{\b B}
&=\fr4L_a{}^b(\g^a{}_b)_\a{}^\b\d_A{}^B+M_A{}^B\delta_\a{}^\b
+\fr2N\d_\a{}^\b\d_A{}^B\komma\cr 
}\eqn    
$$
where $L\in so(1,2)$, $M\in so(8)$ and $N\in{\Bbb R}$.

\subsection\ConventionalConstraints{Conventional 
constraints}As usual
in superspace geometry, the number of fields are reduced using
conventional constraints [\ConvConstr,\CGNT].
The constraints we will impose
are of different types. The property they have in common is that
they are effectuated by fixing some components of the torsion. This
ensures the gauge covariance of the constraints, and therefore of
the resulting physical system. In principle, some of the constraints
have the effect of eliminating certain superfluous components of the
vielbein, \ie\ components that after solving the Bianchi identities occur in
combinations such that they can be removed by field redefinitions
(as can be seen by not enforcing these constraints). However,
imposing them explicitly in terms of vielbeins would not be an 
optimal procedure,
since such constraints could potentially break diffeomorphism invariance. The
vielbeins carry one coordinate index and one inertial index, and the
coordinate index can not be converted into an inertial index (the
result would be the unit matrix). The torsion components, on the
other hand, carry an inertial index and in addition two lower
indices that can be taken in the inertial as well as in the
coordinate basis. All constraints are formulated in terms of the
torsion, and in terms of components with inertial indices only.
Since such components are scalars under diffeomorphisms, this is the
only covariant procedure to impose constraints. As long as they are
formulated in a way that respects the local structure symmetry, all
symmetries will be preserved.

Let us start by considering the conventional constraints. 
There are two kinds of
conventional constraints that can be associated with transformations
of the spin connection and the vielbein respectively, while the
other is held constant. These two transformations have the property
that they leave the torsion Bianchi identities invariant and
therefore take a solution of the Bianchi identities into a new solution. This is
the reason why we can use these kinds of transformations in order to
find an as simple solution to the Bianchi identities as possible. The two kinds
of transformations clearly commute with each other\foot\star{Note,
however, that the commutativity of the constraints 
does not imply that they do not interfere with each other.}.

The first kind shifts the spin connection by an arbitrary 1-form
(with values in the structure algebra) and leaves the vielbein
invariant:
$$
\left.\matrix{\hfill E^\A &\rightarrow& E^\A\hfill\cr
\Omega_\A{}^\B &\rightarrow& \Omega_\A{}^\B+\Delta_\A{}^\B\cr}\right\}
\quad\Longrightarrow\quad T^\A\rightarrow T^\A+E^\B\w\Delta_\B{}^\A\punkt
\Eqn\ConvOmegaTransf
$$
This kind of redefinition serves to remove the independent degrees
of freedom in $\Omega$, which can be achieved by constraints on $T$ as
long as there are no irreducible modules of the structure
group residing in $\Omega$ that do not occur in $T$ (all structure
groups under consideration fulfill this requirement, as will be seen
later). This shift is often expressed as the torsion being absorbed
in the spin connection. The canonical example is ordinary bosonic
geometry, where one gets $T_{ab}{}^c\rightarrow
T_{ab}{}^c+2\Delta_{[ab]}{}^c$, where $\Delta$ is antisymmetric in the last
two indices, meaning that the transformation can be used to set the
torsion identically to zero, leaving the vielbeins as the only
independent variables. In supergravity the analysis is more subtle.
Only certain modules in the torsion can be brought to zero.

The second kind of transformation consists of a change of tangent
bundle, while the connection is left invariant:
$$
\left.\matrix{\hfill E^\A &\rightarrow& E^\B M_\B{}^\A\hfill\cr
\Omega_\A{}^\B &\rightarrow& \Omega_\A{}^\B\hfill\cr}\right\}
\quad\Longrightarrow\quad T^\A\rightarrow T^\B M_\B{}^\A+E^\B\w DM_\B{}^\A\punkt
\Eqn\ConvETransf
$$
Again, it is essential that one implements the constraints on the
torsion. This will mean that not all components in $M$ can be used.
In fact, the remaining degrees of freedom will all reside in the
component $E_\mu{}^a$ of negative dimension. The form of
the transformation of $T$ will in practice mean that the
transformations have to be implemented sequentially in increasing
dimension, in order for the second term not to interfere with
constraints obtained by using the first term. This second kind of
transformation has no relevance in purely bosonic geometry --- there
$M$ has dimension 0, and can not be used to algebraically eliminate
torsion components of dimension 1 (which are taken care of by the
first kind of transformation, anyway). It should also be noted that
not all matrices $M$ are relevant. If $M$ is an element in the
structure group, the transformations in eq.
(\ConvETransf) can be supplemented by a transformation
of the first kind from eq. (\ConvOmegaTransf) with suitable
parameter ($\Delta=M^{-1}dM+M^{-1}\Omega M-\Omega$) so that the total
transformation is a gauge transformation.

\subsection\ImplementationCC{Implementation of the 
constraints}An examination of the transformations of 
the previous subsection at dimension 0 shows that conventional
constraints corresponding to the parameters $M_a{}^b$ and $M_{\a
A}{}^{\b B}$ can be used to eliminate all degrees of freedom in the 
torsion at dimension 0,
except an element in $(4)(0020)$, so that
$$
T_{\a A,\b B}{}^c=2\d_{AB}\g_{\a\b}^c+X_{AB}^{cd}(\g_d)_{\a\b}\komma\eqn
$$
where $X$ is symmetric and traceless in both pairs of indices. This
is what motivates the pure spinor constraints in
section \Multiplets. Dimension zero torsion with field dependent terms like $X$
has been used in other contexts to take the theory off-shell, in
particular for ten-dimensional supergravity 
as discussed in refs. [\BNXXXI,\HoweUmerski]\foot\star{The situation in
ten-dimensional supergravity is special in the sense that the pure
spinor cohomology ({\xit i.e.}, the linearised fields obtained without
this extra tensor) is not entirely off-shell although the theory is
only half-maximally supersymmetric [\NilssonTollsten]. 
No such phenomenon occurs in the
present three-dimensional model.}. 

In addition to the conventional constraints, one needs to make one
additional choice in order to obtain the physical multiplet. We call
this constraint, $X=0$, a physical, or non-conventional, constraint. Setting
$X=0$ is the constraint implied by the $Q$-closedness of
section \Multiplets. This constraint was contained in the definition
of a superconformal structure given in ref. 
[\HoweIzquierdoPapadopoulosTownsend]. Note that the requirement of
super-Weyl invariance 
provides an independent argument for setting $X=0$ to zero 
(see \eg\ ref. [\HoweWeyl]). Super-Weyl transformation are treated in
detail in refs. [\HoweSezginRevisited,\KuzenkoLindstromTartaglino].

At dimension \tf12, one has the constraints corresponding to 
the parameters $M_a{}^{\b B}$, as well as
$(\Delta_{\a A})_a{}^b$, $(\Delta_{\a A})_A{}^B$ and $\Delta_{\a A}$.
It turns out that there is a slight excess of possible
transformations, when Weyl scalings are included in the structure
algebra.
All four parameters contain one spinor $(1)(0010)$ each, but it turns
out that a certain combination of these transformations leaves the
dimension-\tf12\ torsion invariant. 
There are a priori five independent spinors in the torsion at
dimension \tf12, which transforms as
$$
\eqalign{
\delta T_{\a A,b}{}^c&=(\g_b{}^c(\chi_1-2\chi_4))_{\a A}
   +(\chi_3+2\chi_4)_{\a A}\d_b{}^c\komma\cr
\delta T_{\a A,\b B}{}^{\g C}&=
  \e_{\a\b}(\Fr32\chi_1+\fr2\chi_2+\fr2\chi_3)_{[A}^\g\d_{B]}^C
  +\g^i_{\a\b}\d_{AB}(\g_i(\fr2\chi_2+2\chi_4))^{C\g} \cr
  &+\g^i_{\a\b}(\g_i(-\fr2\chi_1-\fr2\chi_2+\fr2\chi_3))_{(A}^\g\d_{B)}^C\komma\cr
}\eqn
$$
where
$$
\eqalign{
(\Delta_{\a A})_b{}^c&=(\g_b{}^c\chi_1)_{\a A}\komma\cr
(\Delta_{\a A})_B{}^C&=
   \fr2(\d_{AB}\chi_{2\a}^C-\d^C_A\chi_{2\a B})\komma\cr
\Delta_{\a A}&=\chi_{3\a A}\komma\cr
M_a{}^{\b B}&=(\g_a\chi_4)^{\b B}\punkt\cr
}\eqn
$$
A shift $\chi_3$ of the Weyl connection can be compensated by
conventional transformations with $\chi_1=-\chi_3$, $\chi_2=2\chi_3$
and $\chi_4=-\fr2\chi_3$. Such a compensating transformation will
affect torsion at higher dimension, due to derivatives in
eq. (\ConvETransf).
We will be especially interested in flat Weyl connections, with gauge
transformations $\d\omega_{\a A}=D_{\a A}\phi$. Such Weyl rescalings
clearly lead to shifts in the dimension 1 torsion with second
spinorial derivatives of $\phi$, \ie, in the modules
$\wedge^2(1)(0010)=(0)(0000)\oplus(0)(0020)\oplus(2)(0100)$. In the
following subsection, this argument is used to gauge away such
unwanted (component) degrees of freedom. In Appendix A, the complete
component expansion of an unconstrained scalar superfield is given.
We should make clear that, unlike in $D=11$ supergravity [\HoweWeyl],
inclusion of scaling in the structure group is not necessary for the
elimination of dimension \tf12 torsion, but used here only to
systematise the form of the scale transformations.

At dimension 1, the shift in the Lorentz spin connection is used to
set $T_{ab}{}^c=0$, as usual.

\subsection\BISolution{Solution of the Bianchi identities}We will 
summarise the results of solving the Bianchi
identities up to dimension 2. In reference
[\KuzenkoLindstromTartaglino], they where solved up to
dimension \tf32\ for arbitrary number of supersymmetries.

After the implementation of the conventional and physical constraints
at dimension 0 and \tf12, we have
$$
\eqalign{
T_{\a A,\b B}{}^c&=2\g^c_{\a\b}\d_{AB}\komma\cr
T_{\a A,b}{}^c&=0\komma\cr
T_{\a A,\b B}{}^{\g C}&=0\punkt\cr
}\eqn
$$ 
At dimension 1, it is straightforward to show that the Bianchi
identities allow the three
modules $(2)(0100)\oplus(0)(0020)\oplus(0)(0000)$
in the torsion. They are introduced as  
$$
T_{a,\b B}{}^{\g C}=\e_{a}{}^{ij}(\g_i)_\b{}^\g Y_{jB}{}^C
+(\g_a)_\b{}^\g(\tZ_B{}^C+\d_B{}^CZ)
\eqn
$$
(in agreement with ref. [\HoweIzquierdoPapadopoulosTownsend]).
The curvature at dimension 1 also contains these superfields.
In addition,
fields in $(0)(2000)\oplus(0)(0002)$ are allowed in the dimension 1
curvature, without appearing in the torsion. This was observed already
in ref. [\HoweIzquierdoPapadopoulosTownsend] based on the above constraint.
This phenomenon is unique to $D=3$, and forbidden by Dragon's theorem
in higher dimensions.
We call it the ``Dragon window''. The appearance of a Dragon window is
connected to the peculiar ``branched'' structure of the supermultiplet
demonstrated in section \Multiplets.
As described in that section, we can 
take the second of these modules, $(0)(0002)$, 
to vanish (while staying off-shell), and denote the first
one $C^+_{ABCD}$, which is completely antisymmetric and selfdual. The
complete curvature at dimension 1 is
$$
\eqalign{
(R_{\a A,\b B})_{cd}&=4\e_{\a\b}\e_{cd}{}^iY_{iAB}
       +4\e_{cd}{}^i(\g_i)_{\a\b}(\tZ_{AB}+\d_{AB}Z)\komma\cr
(F_{\a A,\b B})_{CD}&=-2(\g_i)_{\a\b}\d_{AB}Y^i{}_{CD}
      +8(\g_i)_{\a\b}\d_{[C(A}Y^i{}_{B)D]}\cr
      &-8\e_{\a\b}\d_{[A[C}(\tZ_{D]B]}+\d_{D]B]}Z)
+\e_{\a\b}C^+_{ABCD}\punkt\cr
}\eqn
$$
The Weyl component of the curvature vanishes.
The argument concerning Weyl rescalings in the 
previous subsection shows that the fields $Y$, $\tZ$ and $Z$, although
needed in the off-shell supergeometry, have lowest components that are
pure gauge [\HoweIzquierdoPapadopoulosTownsend] 
(see ref. [\HoweTwoDim] for a similar phenomenon in two
dimensions and the Appendix
for the $\theta$ expansion of a $N=8$ scalar superfield in three dimensions). 
It can of course be checked explicitly that they transform
inhomogeneously under Weyl rescalings, with the compensating
transformations included to keep the dimension \tf12\ torsion
vanishing.

As can be seen  from section \Multiplets, the field $C^+$ seems to be where
conformal couplings are to be inserted\foot\star{Such couplings are
derived in ref. [\GranNilsson] but one should  
note that there are still unresolved questions concerning the exact
structure of such conformal couplings  
in the $\ss N=8$ case. See section 
{\xold4}.{\xold2} for further comments on this
issue.}. This will also be verified 
later by
explicitly checking that setting it to zero yields the equations of
motion of pure conformal supergravity.

Solving the Bianchi identities at dimension \tf32\ shows that the
modules (3)(0110) and (3)(1001) in $DY$, (1)(0030) in
$D\tZ$ and (1)(2010) in $DC^+$ are set to 0 (the latter due to the
curvature Bianchi identity). In addition, the
(1)(1001)'s in $DY$ and $DC^+$ are related as well as the (1)(0110)'s
in $DY$ and $D\tZ$. There is also one linear relation between the
(1)(0010)'s in $DY$, $D\tZ$ and $DZ$. The torsion component (3)(0010)
is identified with the one in $DY$ and the one in (1)(0010) 
with a certain linear
combination of the two surviving (1)(0010)'s. We have not yet bothered
to write down explicit expressions for the curvatures at dimension \tf32, but
they are completely and straightforwardly solved for using the torsion
Bianchi identities. With the concrete Ansatz where every term represents an
irreducible component (the vanishing modules
above are left out)
$$
\eqalign{
D_{\a A}Y_{bCD}&=\d_{A[C}y_{D]b\a}+\d_{A[C}(\g_{|b|}y_{D]})_\a
    +(\g_by_{CD,A})_\a+(\g_by_{ACD})_\a\komma\cr
D_{\a A}\tZ_{BC}&=\d_{A(B}\tz_{C)'\a}+\tz_{A(B,C)\a}\komma\cr
D_{\a A}Z&=z_{\a A}\komma\cr
D_{\a A}C^+_{BCDE}&=4\d_{A[B}c_{CDE]\a}+\fr6\e_{ABCDE}{}^{FGH}c_{FGH\a}\komma\cr
T_{ab}{}^{\g C}&=\e_{ab}{}^i(\tilde t_i+\g_it)^{\g C}\cr
}\Eqn\DimThreeHalvesAnsatz
$$
(primed symmetrisation includes subtraction of the trace),
the relations become
$$
\eqalign{
(1)(1001):\;0&=y_{ABC\a}+\fr6c_{ABC\a}\komma\cr
(1)(0110):\;0&=2y_{AB,C\a}+\tz_{AB,C\a}\komma\cr
(1)(0010):\;0&=\Fr32y_{A\a}-\fr4\tz_{A\a}-2z_{A\a}\komma\cr
}\eqn
$$
and the torsion is
$$
\eqalign{
\tilde t_{a\a A}&=\fr2y_{a\a A}\komma\cr
t_{\a A}&=\fr2y_{\a A}+\fr2\tz_{\a A}\punkt\cr
}\Eqn\ThreeHalfTorsion
$$
Also some of the remaining component fields at dimension \tf32\ can be
removed by a Weyl rescaling, since they come at the $\theta^3$ level
of a scalar superfield. This applies to one of the spinors $(1)(0010)$
and the field in $(1)(0110)$. This leaves at dimension \tf32\
precisely the torsion modules together with the field in $(1)(1001)$
in the current multiplet.

Now to dimension 2. 
The Bianchi identity with fermionic indices is
$$
(R_{ab})_{\g C}{}^{\d D}=2D_{[a}T_{b]\g C}{}^{\d D}+D_{\g
C}T_{ab}{}^{\d D}+2T_{[a|\g C}{}^{\e E}T_{b]\e E}{}^{\d D}\punkt
\Eqn\DimTwoBI
$$
It is {\it a priori} not clear to what
extent the torsion Bianchi identities 
should give new information here, in addition
to the one obtained by acting with fermionic derivatives on the
relations on dimension \tf32. 

We will go through the modules appearing at dim. 2 systematically, and
show that all essential information off-shell is provided by the
equations at dimension \tf32. The analysis will also provide
information concerning the absence of equations of motion at dimension
2, the appearance of the physical curvature and some implications of
going on a conformal shell by specifying $C^+$.

\underbar{\bf(0)(0020):}
The module (0)(0020) at dim. 2 comes from (1)(0010),
(1)(0110) and (1)(0030) at dim. \tf32. One must therefore use the
identities $D\tZ|_{(1)(0030)}=0$, \ie, 
$D_{\a(A}\tZ_{BC)}-\fr5\d_{(AB}D_{|\a|}{}^D\tZ_{C)D}=0$,
 together with
$2y_{AB,C\a}+\tz_{AB,C\a}=0$, and of course the Bianchi identity. 
We act on the former equation with one more fermionic derivative and
contract the spinor index and one of the $SO(8)$ indices to get
(0)(0020). In the process, we have to divide the two fermionic
derivatives in symmetric and antisymmetric parts, of which the
symmetric part gets contribution from the curvature at dim. 1 (and in
general from torsion, but not in this case). This gives a relation
between the two (0)(0020)'s in $D^{\wedge 2}\tZ$:
$$
0=\left\{(D_{(A}D_{C)}){\tZ}_B{}^C+\Fr58(D^CD_C)\tZ_{AB}
    -24(\tZ+Z\id)^2_{AB}\right\}_{(AB)'}\komma \Eqn\TZreps
$$ 
where the overall traceless symmetrisation in $(AB)'$ is taken after
the other ones. The derivative of the relation
$2y_{AB,C\a}+\tz_{AB,C\a}=0$ is treated in the same way. Taking one
derivative of the identity (the inversion of how $y_{AB,C\a}$ appears
in the Ansatz (\DimThreeHalvesAnsatz)) 
$$
y_{AB,C\a}=\fr3(\g_iD)_{\a C}Y^i{}_{AB}-\fr3(\g_iD)_{\a
[C}Y^i{}_{AB]} +\Fr2{21}\d_{C[A}(\g^iD)_{|\a}^EY_{i|B]E}\eqn
$$ 
gives
$$
(D^Ay_{A(B,C)})=\left\{\Fr8{21}(D_{[B}\g^iD_{E]})Y_{iC}{}^E
-\Fr{32}7(Y^iY_i)_{BC}\right\}_{(BC)'}\komma\eqn
$$
and similarly on $\tZ$, where also eq. (\TZreps) has been used,
$$
(D^A\tz_{A(B,C)})=\Fr87(D^ED_E)\tZ_{BC}-\Fr{64}7(\tZ+Z\id)^2_{(BC)'}\punkt\eqn
$$
Combining this information, we have, already before using the Bianchi
identity at dim. 2, 
$$
0=\left\{2(D_{[A}\g^iD_{E]})Y_{iB}{}^E-24(Y^iY_i)_{AB}
     +3(D^ED_E)\tZ_{AB}-24(\tZ+Z\id)^2_{AB}\right\}_{(AB)'}\Eqn\DimTwoRel
$$
To obtain similar information from the Bianchi identity, 
we need to perform a similar
calculation obtaining (0)(0020) as a fermionic derivative on the
fields in (1)(0010), since they sit in the torsion. Using the ``inversions''
$y_{\a A}=-\Fr2{21}(\g^iD)_\a^EY_{iAE}$ and $\tz_{\a
A}=\Fr8{35}D_\a^E\tZ_{AE}$, one obtains
$$
\eqalign{
(D_{(A}y_{B)'})&=\left\{-\Fr2{21}(D_{[A}\g^iD_{E]})Y_{iB}{}^E
             -\Fr{32}{21}(Y^iY_i)_{AB}\right\}_{(AB)'}\komma\cr
(D_{(A}\tz_{B)'})&=-\fr7(D^ED_E)\tZ_{AB}+\Fr{64}7(\tZ+Z\id)^2_{(AB)'}\punkt\cr
}\eqn
$$
These equations are then used in the (0)(0020) component of the
Bianchi identity, using the form of the lower-dimensional torsion from
above. The result turns out to be exactly eq. (\DimTwoRel) again
(after an intricate combination of a number of numerical
coefficients), 
so no new
information is obtained. We consider this quite striking result 
a check that the calculation
is correct. 
This means that there is one unrestricted (0)(0020) field left at
dim. 2.

The (0)(0020) in $Z$ can of
course also be solved for by using the relation
$\Fr32y_{A\a}-\fr4\tz_{A\a}-2z_{A\a}=0$. It becomes
$$
\eqalign{
&(D_{(A}D_{B)'})Z\cr
&=-\fr{56}\left\{4(D_{[A}\g^iD_{E]})Y_{iB}{}^E+64(Y^iY_i)_{AB}
     -(D^ED_E)\tZ_{AB}+64(\tZ+Z\id)^2_{AB}\right\}_{(AB)'}\cr
&=-\left\{\fr{12}(D_{[A}\g^iD_{E]})Y_{iB}{}^E+(Y^iY_i)_{AB}
     +(\tZ+Z\id)^2_{AB}\right\}_{(AB)'}\cr
}\Eqn\ZDimTwo
$$

\underbar{\bf(2)(0100):}
The module (2)(0100) is interesting, since this is where the
$SO(8)$ field strength occurs. {\it A priori} there are 3 (2)(0100)'s
in $D^{\wedge 2}Y$ and one each in $D^{\wedge 2}\tZ$, $D^{\wedge 2}Z$ and
$D^{\wedge 2}C^+$. We will use the shorthand
$$
\eqalign{
Y^{(2)}_1&=(D^ED_E)Y_{aAB}\komma\cr
Y^{(2)}_2&=\left\{(D_{(A}D_{E)})Y_{aB}{}^E\right\}_{[AB]}\komma\cr
Y^{(2)}_3&=\left\{\e_{aij}(D_{[A}\g^iD_{E]})Y^j{}_B{}^E\right\}_{[AB]}\komma\cr
\tZ^{(2)}&=\left\{(D_{[A}\g_aD_{E]})\tZ_B{}^E\right\}_{[AB]}\komma\cr
Z^{(2)}&=(D_{[A}\g_{|a|}D_{B]})Z\komma\cr
C^{+(2)}&=(D^C\g_aD^D)C^+_{ABCD}\komma\cr
\e F&=\e_a{}^{ij}(F_{ij})_{AB}\punkt\cr
}\eqn
$$
We will for the moment not deal with $Z^{(2)}$ and $C^{+(2)}$. The
dimension \tf32 torsion does not contain $DC^+$ and can be written
without $DZ$. 
We also use shorthand for the other structures appearing:
$$
\eqalign{
DY&=\e_a{}^{ij}D_iY_{jAB}\komma\cr
YY&=\e_a{}^{ij}(Y_iY_j)_{AB}\komma\cr
\tZ Y&=(\tZ Y_a)_{[AB]}\komma\cr
ZY&=ZY_{aAB}\komma\cr
C^+Y&=C^+_{AB}{}^{CD}Y_{aCD}\punkt\cr
}
$$

The three structures in $Y$ have two relation among themselves (and
lower-dimensional fields) thanks to the vanishing of the
modules (3)(0110) and (3)(1001) in $DY$. The equations at
dim. \tf32 read (the (3)(0110) equation as stated also contains (3)(1001))
$$
\eqalign{
(3)(1001):\;0&=D_{\a[A}Y_{|a|BC]}-\fr3(\g_a\g^iD)_{\a[A}Y_{|i|BC]}\komma\cr
(3)(0110):\;0&=D_{\a C}Y_{aAB}+\Fr27\d_{C[A}D^E_{|\a}Y_{a|B]E}\cr
     &-\fr3(\g_a\g^iD)_{\a C}Y_{iAB}
      -\Fr2{21}\d_{C[A}(\g_{|a}\g^iD)^E_\a Y_{i|B]E}\punkt\cr
}\eqn
$$
Acting on these equation with one further fermionic derivative,
forming (2)(0100), separating into antisymmetric and symmetric
products of derivatives (the former leading to the $Y^{(2)}$'s, the
latter to torsion and curvature) leads to the equations
$$
\eqalign{
0&=Y^{(2)}_1+2Y^{(2)}_2-Y^{(2)}_3-6DY+8YY+8\tZ Y+72ZY-2C^+Y\komma\cr
0&=14Y^{(2)}_1+4Y^{(2)}_2+2Y^{(2)}_3-108DY+144YY-48\tZ Y+336ZY+4C^+Y
\punkt\cr
}\eqn
$$
As for the module (0)(0020), a relation at dim. 2 is obtained
by applying a fermionic derivative on the (1)(0110) relation
$2y_{AB,C\a}+\tz_{AB,C\a}=0$, and forming (2)(0100). This results in
$$
0=7Y^{(2)}_1-4Y^{(2)}_2-10Y^{(2)}_3-24\tZ^{(2)}
         +120DY-40YY-400\tZ Y-720ZY+10C^+Y\punkt\eqn
$$
At this point, there is one independent field left. The (2)(0100)'s in
$Z$ and $C^+$ are related to the ones in $Y$ and $\tZ$ via (1)(0010)
and (1)(1001) respectively, but they are not needed in the torsion
Bianchi identity.

The torsion Bianchi identity contains two (2)(0100)'s. One 
of these is obtained as
$\e_{a}{}^{ij}\e^{\a\b}(R_{ij})_{\a[A,|\b|B]}=\ldots$, and contains $F$; the
other one as $(\g^i)^{\a\b}(R_{ai})_{\a[A,|\b|B]}=\ldots$, and does
not contain curvature. The latter one can potentially give one more
relation among the fields. Some calculation gives the equation
$$
0=5Y^{(2)}_3+8\tZ^{(2)}
         -60DY+40YY+112\tZ Y+240ZY\punkt\eqn
$$
This equation is a linear combination of the three earlier ones. The
other equation in the Bianchi identity expresses $F$ in terms of the remaining
field. 
The result is
$$
\e F=\fr7Y^{(2)}_2-\Fr4{35}\tZ^{(2)}
         -\Fr{10}7YY+\Fr{24}{35}\tZ Y-\Fr{24}7ZY+\fr7C^+Y\punkt\Eqn\FExpression
$$

Setting $C^+=0$ implies going on shell in pure conformal
supergravity. This constraint leads to additional relations at
dim. 2, since it then follows that $DY|_{(1)(1001)}=0$. Taking another
fermionic derivative on this equation and forming (2)(0100) adds one
more relation between the fields in this module at dim. 2,
which means that there is no free components left. The extra equation
is
$$
0=Y^{(2)}_1+2Y^{(2)}_2+2Y^{(2)}_3
         +12DY-40YY+32\tZ Y\punkt\eqn
$$
Solving for the (2)(0100) components of the superfields and inserting
in eq. (\FExpression) gives the very nice equation
$$
(F_{ab})_{CD}=-2D_{[a}Y_{b]CD}+2(Y_{[a}Y_{b]})_{CD}\eqn
$$
(the terms with $\tZ Y$ and $ZY$ cancel out), or equivalently as
forms, $F=-DY-Y\wedge Y$.
It should be noted that this expression, for this specific combination
of numerical coefficients, automatically satisfies the
Bianchi identity for $F$. It can be written in the more suggestive way
$$
F(A+Y)=0\komma\Eqn\FExpressionTwo
$$
where $A$ is the $SO(8)$ gauge potential,
stating that $A+Y$ is a flat connection. We remind that the lowest
component of $Y$ is gauged away by a Weyl transformation.
The only other effect of setting $C^+=0$ (or equal to some current
superfield) is that the auxiliary field in (0)(0002) at dim. 2, which
is unconstrained by the Bianchi identities, will vanish.

More generally, $Z^{(2)}_{aAB}$ and $C^{+(2)}_{aAB}$ may be expressed
as
$$
\eqalign{
Z^{(2)}&=\fr{70}(-5Y^{(2)}_2+5Y^{(2)}_3-2\tZ^{(2)}
         +10DY+20YY+32\tZ Y+80ZY-5C^+Y)\komma\cr
C^{+(2)}&=\Fr{10}3(Y^{(2)}_1+2Y^{(2)}_2+2Y^{(2)}_3
         +12DY-40YY+32\tZ Y-2C^+Y)\punkt\cr
}\eqn
$$
This allows for reexpressing the field strength as
$$
\e_a{}^{bc}F(A+Y)_{bcAB}=\fr{60}C^{+(2)}_{aAB}+\fr3C^+_{AB}{}^{CD}Y_{aCD}\punkt\eqn
$$

\underbar{\bf (0)(0100):} There are from the beginning two (0)(0100)'s, one each in
$D^2Y$ and $D^2\tZ$. The only other term that can enter is
$D^iY_{iAB}$. The constraints in (1)(0010), (1)(1001) and (1)(0110)
propagate to this module. 
There is also a Bianchi identity in this module.
We define $Y^{(2)}_{AB}=\{(D_{[A}\g^iD_{C]})Y_{iB}{}^C\}_{[AB]}$ and 
$\tZ^{(2)}_{AB}=\{(D_{(A}D_{C)})\tZ_B{}^C\}_{[AB]}$. The (1)(1001)
constraint gives $Y^{(2)}_{AB}+6D^iY_{iAB}=0$. The (1)(0010) and
(1)(0110) constraints give respectively
$Y^{(2)}_{AB}+2D^iY_{iAB}+\Fr25\tZ^{(2)}_{AB}=0$ and 
$Y^{(2)}_{AB}-12D^iY_{iAB}+\Fr95\tZ^{(2)}_{AB}=0$. There is a linear
dependency, and the solutions are $Y^{(2)}_{AB}=-6D^iY_{iAB}$,
$\tZ^{(2)}_{AB}=10D^iY_{iAB}$. 
The Bianchi identity, finally, gives 
$Y^{(2)}_{AB}+30D^iY_{iAB}-\Fr{12}5\tZ^{(2)}_{AB}=0$, which provides
no further information. 

\underbar{\bf(2)(0000):} Similarly, there is one (2)(0000) in $D^2Y$. 
It becomes constrained by
the (1)(0010) constraint and by two (2)(0000)'s in the Bianchi
identity (neither of
which contain curvature, due to $(R_{[ab})_{c]}{}^d=0$). The only
other expression in this module is $D_aZ$. 
The components of the Bianchi identity are obtained as $(R_{ab})_{\g C}{}^{\g
C}=\ldots$ and $(R_{ab})_{\g C}{}^{\d C}(\g^b)_\d{}^\g=\ldots$. The
former gets contribution only from the second term on the right hand
side of eq. (\DimTwoBI), and is trivially fulfilled due to the easily
derived equation $(D^Ey_{aE})+(D^E\g_ay_E)=0$. The second one gives
$\e_a{}^{ij}(D^E\g_iD^F)Y_{jEF}=-224D_aZ$. Finally, one gets an
equation in (2)(0000) by differentiation of the (1)(0010) equation 
$\Fr32y_{A\a}-\fr4\tz_{A\a}-2z_{A\a}$, with contribution from the
first and third terms. It turns out to yield the same information
again. Thus, there is no restriction on $D_aZ$, and no remaining
degrees of freedom in (2)(0000) at dim. 2.

\underbar{\bf(2)(0020):} In (2)(0020), there are two fields in $D^2Y$
and one in $D^2\tZ$. The 
constraints in (1)(0010), (1)(0030), (1)(0110) and (3)(0110) propagate
to this module, and there are two (2)(0020)'s in the Bianchi identity. This
looks dangerous; there is na\"\i vely 6 equations for 3 fields. The
equations may also contain $D_a\tZ_{AB}$ and $(Y_a\tZ)_{(AB)}$. Let
$$
\eqalign{
Y^{(2)}_{1\,aAB}&=\{(D_{(A}D_{E)})Y_{aB}{}^E\}_{(AB)}\komma\cr
Y^{(2)}_{2\,aAB}&=\{\e_a{}^{ij}(D_{[A}\g_iD_{E]})Y_{jB}{}^E\}_{(AB)'}\komma\cr
\tZ^{(2)}_{aAB}&=\{(D_{[A}\g_aD_{E]})\tZ_B{}^E\}_{(AB)}\punkt\cr
}\eqn
$$
Beginning with the constraint $DY|_{(3)(0110)}=0$, it gives
$3Y^{(2)}_{1\,aAB}-2Y^{(2)}_{2\,aAB}-36(Y_a\tZ)_{(AB)}=0$.
The constraint $D\tZ|_{(1)(0030)}=0$ propagates to
$\tZ^{(2)}_{aAB}+8D_a\tZ_{AB}+4(Y_a\tZ)_{(AB)}=0$. The constraints
in (1)(0110) and (1)(0010) give the two equations
$Y^{(2)}_{1\,aAB}+\Fr43Y^{(2)}_{2\,aAB}-\tZ^{(2)}_{aAB}
+16D_a\tZ_{AB}-64(Y_a\tZ)_{(AB)}=0$ and
$-5Y^{(2)}_{1\,aAB}+5Y^{(2)}_{2\,aAB}-2\tZ^{(2)}_{aAB}
+4D_a\tZ_{AB}+12(Y_a\tZ)_{(AB)}=0$. Finally, the two equations from
the Bianchi identity are
$5Y^{(2)}_{1\,aAB}-4\tZ^{(2)}_{aAB}
+8D_a\tZ_{AB}-156(Y_a\tZ)_{(AB)}=0$ and
$5Y^{(2)}_{2\,aAB}+8\tZ^{(2)}_{aAB}
+124D_a\tZ_{AB}-88(Y_a\tZ)_{(AB)}=0$. Only three of the six equations
are linearly independent, and the solutions are
$$
\eqalign{
Y^{(2)}_{1\,aAB}&=-8D_a\tZ_{AB}+28(Y_a\tZ)_{(AB)}\komma\cr
Y^{(2)}_{2\,aAB}&=-12D_a\tZ_{AB}+24(Y_a\tZ)_{(AB)}\komma\cr
\tZ^{(2)}_{aAB}&=-8D_a\tZ_{AB}-4(Y_a\tZ)_{(AB)}\punkt\cr
}\eqn
$$

\underbar{\bf(0)(0000):} Concerning the singlet at dim. 2, we have
initially one field each in 
$Y$, $\tZ$ and $Z$. One relation is obtained from the (1)(0010)
identity at dim. \tf32, and reads
$$
0=\fr7Y^{(2)}+\Fr2{35}\tZ^{(2)}+2Z^{(2)}\komma\eqn
$$
where $Y^{(2)}=(D^A\g^iD^B)Y_{iAB}$, $\tZ^{(2)}=(D^AD^B)\tZ_{AB}$,
$Z^{(2)}=(D^ED_E)Z$. The (0)(0000) part of the Bianchi identity
determines the scalar curvature as
$$
R=-\fr{14}Y^{(2)}+\Fr6{35}\tZ^{(2)}+2\tr(Y^iY_i)-6\tr(\tZ+Z\id)^2\punkt\eqn
$$

\underbar{\bf(4)(0000):} In (4)(0000), the Bianchi identity 
relates the traceless
part of the Ricci tensor to 
the field $Y^{(2)}_{ab}=(D^E\g_{(a}D^F)Y_{b)'EF}$ as
$$
R_{(ab)'}=-\fr{56}Y^{(2)}_{ab}+\fr2\tr(Y_{(a}Y_{b)'})\punkt\eqn
$$

The last components of the Bianchi identity lie in (4)(0100) and (4)(0020). The
fields in $D^2Y$ in these modules (one each) must become determined
in terms of $D_{(a}Y_{b)'AB}$ and $(Y_{(a}Y_{b)'})_{AB}$,
respectively, by the vanishing of the (3)(1001) and (3)(0110) parts of
$DY$. The Bianchi identities must then constitute consistency checks on those
relations. We have not yet checked this.

Of the surviving components at dimension 2, some can be removed by a
Weyl transformation. This applies to one of the scalars $(0)(0000)$
and the modules $(0)(0020)$ and $(0)(0200)$. 

\vskip\baselineskip
\boxit{
\vskip.5\baselineskip
\centerline{\epsffile{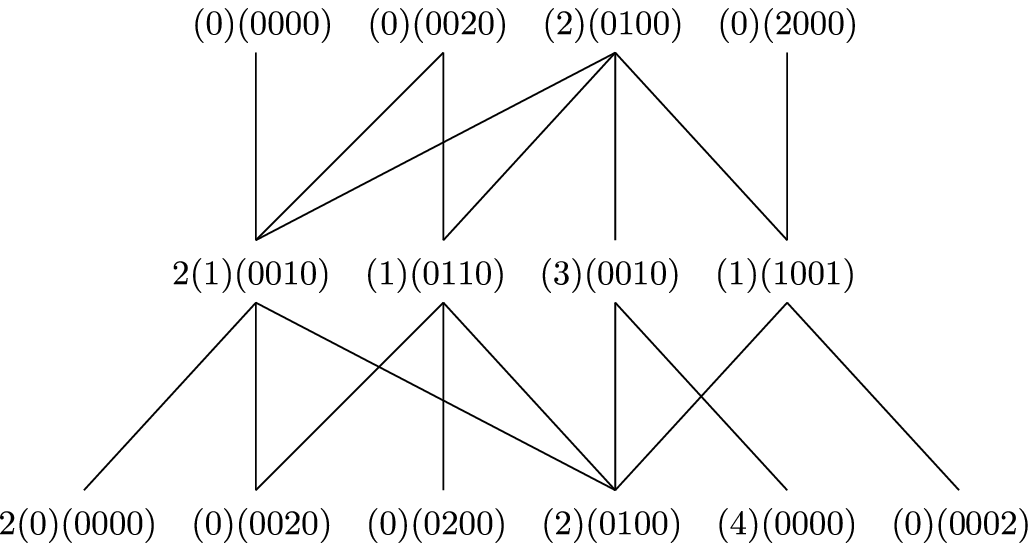}}
\vskip\baselineskip
}

\noindent The above figure is obtained by subtracting all
``descendants'' from the dimension \tf32 constraint. 
Continued subtraction at higher dimensions indicates that no fields
arise at dimension \tf52 or higher.

\section\MatterCoupling{Equations of motion and matter couplings}
\immediatesubsection\ConformalEOM{Conformal equations of motion}In $D=3$, 
the Riemann
tensor is completely determined by the Ricci tensor:
$$
R_{ab,cd}=\e_{ab}{}^i\e_{cd}{}^j(R'_{ij}-\fr6\eta_{ij}R)\komma\eqn
$$
where $R'_{ab}=R_{(ab)'}$ is the traceless part of $R_{ab}$. The curvature Bianchi identity is here a vector equation
(after dualisation). It reads
$$
D^bR'_{ab}-\fr6D_aR=0\punkt\eqn
$$
This equations eliminates the curvature scalar as local degree
of freedom. It also coincides with the vector part of the
conformal equation of motion $D_{[a}(R_{b]c}-\fr4\eta_{b]c}R)=0$, 
whose only other content is given by the Cotton tensor in
$(4)(0000)$. The linearised equations of motion for pure conformal
supergravity read.
$$
\e^{\ms}_{(a}{}^{ij}D^{\ms}_{|i}R'_{j|b)'}=0\punkt\Eqn\CottonEq
$$

The Bianchi identity for the gravitino at dim. \tf52 is 
$D^i\tilde t_i+\Dslash t=0$. The conformal equation of motion in
$(3)(0010)$ for the gravitino is 
$$
\e_a{}^{ij}D_i\tilde t_{j\a A}-(\g\hbox{-trace})=0\punkt\Eqn\CottinoEq
$$

We will now show that going on the conformal shell by specifying $C^+$
gives these last two equations. In doing this we will not deal with any specific type of
matter couplings, only pure conformal supergravity, obtained from
$C^+=0$. We will also restrict ourselves to a linearised treatment.

The lowest-dimensional conformal equation of motion is the one for the
$so(8)$ field strength in $(2)(0100)$, obtained in the previous
section. We therefore take a fermionic derivative on the equations in
$(2)(0100)$ listed in section \BianchiIdentities, and project on
$(3)(0010)$. A priori, $Y$ contains 3 components in $(3)(0010)$ at
$\theta^3$, while $\tZ$ and $Z$ contain none. This means, that these
three components are determined in terms of lower-dimensional fields
by taking one spinorial derivative on the three equations in
$(2)(0100)$ at dimension 2. If in addition $C^+=0$, this gives one
more equation, so that $Z^{(2)}$ and $\tZ^{(2)}$ in $(2)(0100)$ can be
solved for. Linearised, they become proportional to
$\e_a{}^{ij}D_iY_{jAB}$. Taking another spinorial derivative and
projecting on $(3)(0010)$ directly gives the gravitino equation of
motion (\CottinoEq).

Acting with one more spinorial derivative and projecting on the module
$(4)(0000)$, straightforwardly gives the Cotton equation (\CottonEq).

We have focused on deriving the equations of motion of conformal
supergravity from the off-shell multiplet, and the connection between
this question and the occurrence of the Dragon window. Obviously, it
is also possible to go on a Poincar\'e shell. The corresponding physical
torsion constraints are clearly present: the tensor $Y$ should then be
identified with the $SO(8)$ current, which will introduce a
dimensionally correct coupling. We do not work out the details of this
construction here but the reader may consult ref.
[\KuzenkoLindstromTartaglino] 
for more information on this issue.

\subsection\BLGCoupling{Matter coupling}A specific 
candidate for coupling to the conformal supergravity is the
$N=8$ scalar multiplet, \ie, the fields of the BLG model. This
coupling has been (partly) constructed in a component formalism
[\GranNilsson], where, however, there are some issues with the closure
of the supersymmetry algebra. 
It is clear from above that the scalar superfield of the BLG model 
[\CederwallBLG,\CederwallABJM] can be used as a source for the
conformal supergravity. The superfield has dimension \tf12\ and
transforms as $(0)(1000)$ (and
carries an additional internal 3-algebra index). A natural choice is
$$
C^{+IJ}=\tr(\Phi^I\Phi^J)\komma\Eqn\MatterCurrent
$$
the consistency of which is shown as follows. 
The constraint on the scalar superfield is
$$
D_{\a A}\Phi^I=(\sigma^I\Psi)_{\a A}\komma\eqn
$$
where $\Psi$ is the fermion superfield, \ie, the module $(1)(1010)$ in
$D\Phi$ vanishes. This implies that the current superfield $C^+$
formed as eq. (\MatterCurrent) fulfills the Bianchi identity
$DC^+|_{(1)(2010)}=0$, which is the only constraint on $C^+$.
A related discussion can be found in ref. [\HoweSezginRevisited].

At the same time as matter sources the supergravity, one should
formulate the matter dynamics in a conformal supergravity
background. This may be achieved if one shows that a nilpotent BRST operator
$Q=\l^{\a A}D_{\a A}$ can be constructed, where $D_{\a A}$ carries the
geometric information. Since $\Phi^I$ transforms under $so(8)$, the
nilpotency of $Q$
may potentially be ruined by $so(8)$ curvature at dimension 1. One
also has to take into consideration that $\Phi^I$ has an extra gauge
invariance in addition to the pure spinor constraint, 
$\Phi^I\equiv\Phi^I+(\l\sigma^I\varrho)$, for arbitrary $\varrho$ in 
$(1)(0001)$. $Q^2\Phi^I$ needs to vanish only modulo such a term.
This gauge invariance turns out to save the nilpotency for the
contributions to the $so(8)$ curvature from $Y$, $\tZ$ and $Z$, but
not from $C^+$. The latter curvature component gives
$$
Q^2\Phi^I\sim\e_{\a\b}\l^{\a A}\l^{\b
B}C^+_{ABCD}(\sigma^{IJ})^{CD}\Phi^J \komma\eqn
$$
which does not take the form of a gauge transformation. 
Expressing $C^+$ as $\Phi^2$ does not change this
fact. This looks like an obstruction to a consistent coupling
between the BLG model and conformal supergravity, which
seems to indicate problems to close the supersymmetry
algebra perhaps related to similar  unresolved problems in
ref. [\GranNilsson]. 
We feel, however, that a proper supersymmetric treatment of the interacting
gravity--matter system calls for a full Batalin--Vilkovisky
formulation, and that only BRST is not enough beyond the linearised
level. Such a formulation so far exists only for the BLG sector
[\CederwallBLG], but experience from eleven dimensions 
[\PureSG] gives reason to
hope that it is achievable.

\vfill\eject
\section\Conclusions{Conclusions and comments}Starting from the $N=8$
superspace BI's in three dimensions we have in 
this paper extended the analysis of
ref. [\HoweIzquierdoPapadopoulosTownsend]  and 
shown that 
we can generate either Poincar\'e or conformal
supergravity. The field equations are ``off-shell'' in the sense that
they  
appear with currents that can be chosen arbitrarily opening up for
constructions containing   supergravity theories, either Poincar\'e or
conformal, coupled to matter multiplets like the BLG theory. One
interesting result is the so called ``Dragon window'', which refers to
the fact that some modules in the supercurvature
tensor manage to avoid Dragon's theorem. This means that in the
analysis of the BI's it is no longer sufficient to solve only the
torsion BI's, since these do not imply all of the curvature
BI's. According to Dragon's theorem this will not happen in four and
higher dimensions. This phenomenon was also used
in refs. [\HoweIzquierdoPapadopoulosTownsend,\HoweSezginRevisited,\KuzenkoLindstromTartaglino]. 

One of the motivations for this work was in fact to get a better
understanding of the topologically gauged BLG theory presented
in ref. [\GranNilsson]. 
The construction in this work exactly corresponds to the coupling
discussed in section \BLGCoupling, 
namely between $N=8$ superconformal gravity in three
dimensions and the $N=8$ BLG theory. However, in ref. [\GranNilsson] 
the attempt to
derive the complete Lagrangian met with some difficulties half way
through the construction for reasons that were not completely
understood. It would therefore be of some value to have an alternative
method by which this theory could be derived. Hopefully the results
obtained here will eventually prove to provide such a method. 

It should be mentioned also that the corresponding construction in the
$N=6$ case, then involving the ABJM theory for $N$-stacks of M2-branes,
was successfully carried out in ref. [\ChuNilsson] following the same steps
and ideas as 
applied in ref. [\GranNilsson] in the $N=8$ topologically gauged BLG
case. For the topologically gauged ABJM theory some complications did
arise but of a much milder kind than for $N=8$ and could be solved. 

During the completion of this paper, related work appeared [\GreitzHowe] which
analyses the case with maximal supersymmetry in three dimensions, 
\ie, $N=16$, and hence does not address the same M2 related questions as
in this paper. Nevertheless, this work solves parts of the three
dimensional BI's and does contain calculations that partly overlap
with ours. 

\appendix{The field content in a scalar superfield}Weyl
rescalings are performed with a scalar superfield of dimension 0. In order to
understand which component fields may be removed by such a rescaling,
its component expansion must be examined. The Dynkin labels of the
modules $R_n$ appearing at order $\theta^n$
are given below. We only list the fields up to
$\theta^8$, since the modules appearing at $\theta^n$ and
$\theta^{16-n}$ are the same.

\vskip3\parskip
\thicksize=\thinsize
\ruledtable
$n$|$R_n$\cr
$0$|$(0)(0000)$\cr
$1$|$(1)(0010)$\cr
$2$|$(0)(0000)\oplus(0)(0020)\oplus(2)(0100)$\cr
$3$|$(1)(0010)\oplus(1)(0110)\oplus(3)(1001)$\cr
$4$|$R_2\oplus(0)(0200)\oplus(2)(1011)\oplus(4)(0002)\oplus(4)(2000)$\cr
$5$|$R_3\oplus(1)(1101)\oplus(3)(0012)\oplus(3)(2010)\oplus(5)(1001)$\cr
$6$|$R_4\oplus(0)(2002)\oplus(2)(0102)\oplus(2)(2100)\oplus(4)(1011)\oplus(6)(0100)$\cr
$7$|$R_5\oplus(1)(1003)\oplus(1)(3001)\oplus(3)(1101)\oplus(5)(0110)\oplus(7)(0010)$\cr
$8$|$R_6\oplus(0)(0004)\oplus(0)(4000)\oplus(2)(2002)\oplus(4)(0200)\oplus(6)(0020)\oplus(8)(0000)$
\endruledtable
\vskip2\parskip

\refout

\end